\documentclass{article}

\usepackage{arxiv}

\usepackage[utf8]{inputenc} 
\usepackage[nottoc]{tocbibind}
\usepackage[T1]{fontenc}    
\usepackage{hyperref}       
\usepackage{url}            
\usepackage{booktabs}       
\usepackage{amsfonts}       
\usepackage{nicefrac}       
\usepackage{microtype}      
\usepackage{cite}
\usepackage{amsmath,amssymb,amsfonts}
\usepackage{algorithmic}
\usepackage{graphicx}
\usepackage{textcomp}
\usepackage{xcolor}
\usepackage{hyperref}
\def\BibTeX{{\rm B\kern-.05em{\sc i\kern-.025em b}\kern-.08em
    T\kern-.1667em\lower.7ex\hbox{E}\kern-.125emX}}
\usepackage{paralist}

\title{A Tertiary and Secondary Study Canvas}

\author{
Bianca M. Napole\~ao\\
Laboratoire d'informatique formelle\\
Université du Québec à Chicoutimi \\
555, boulevard de l'Université, G7H 2B1 \\
Chicoutimi, QC, Canada \\
\texttt{bianca-minetto.napoleao1@uqca.ca} 
\And
Fabio Petrillo\\
Laboratoire d'informatique formelle\\
Université du Québec à Chicoutimi\\
555, boulevard de l'Université, G7H 2B1\\
Chicoutimi, QC, Canada\\
\texttt{fabio@petrillo.com} 
\And
Sylvain Hall\'e\\
Laboratoire d'informatique formelle\\
Université du Québec à Chicoutimi \\
555, boulevard de l'Université, G7H 2B1 \\
Chicoutimi, QC, Canada \\
\texttt{shalle@acm.org} 
}

\begin{document}
\maketitle

\begin{abstract}
Over the past years, more secondary (Systematic Literature Reviews and Systematic Mappings) and tertiary studies have been conducted. Their conduction is considered a quite large task and labor-intensive since it involves a detailed process including a protocol development, which is one of the most challenging phase reported by the software engineering research community. In this scenario, we propose a Secondary and Tertiary Study Canvas aiming to simplify and clarify the understanding of the steps that need to be performed during the secondary and tertiary process conduction, including the protocol development. For this, we synthesized and organized the existing secondary studies' protocols in a Canvas format as well as suggesting a step-based approach to assist the secondary and tertiary studies' conduction. 
\end{abstract}


\section{Introduction}
\label{sec:introduction}

Secondary studies (Systematic Literature Review (SLR) and Systematic Mapping (SM)) are considered a key element of the Evidence-based Software Engineering (EBSE) since their objective is to synthesize evidence regarding a specific topic of interest providing a complete and fair evaluation of the state-of-art about a given  topic\cite{Kitchenham15}. 

In the literature, there are well-established guidelines to conduct secondary and tertiary studies. They are: \textit{Kitchenham \& Charters} Guidelines \cite{Kitchenham07} for secondary and tertiary studies and \textit{Petersen et al.} \cite{Petersen15} which is specific for SM. In 2015, \textit{Kitchenham et al.} published a book \cite{Kitchenham15} addressing a revision of secondary and tertiary studies process and procedures.  \textit{Kitchenham et al.} \cite{Kitchenham10a} pointed out some differences between SLR and SM regarding  the research questions objective, search strings, quantity of selected studies, data extraction and synthesis. Despite these differences, the process of conducting SLRs and SMs is the same. It is important to highlight that \textit{Petersen et al.} \cite{Petersen15} followed \textit{Kitchenham \& Charters} Guidelines \cite{Kitchenham07} to conduct their guideline study. Besides, according to the study conducted by \textit{Napole\~ao et al.}
\cite{Napoleao17}, in practice only the quality assessment of candidate studies is conducted differently between SLRs and SMs. 

Since the introduction of EBSE in the Software Engineering (SE) area in 2004, more and more secondary and tertiary studies have been conducted \cite{Zhang11}.
One of the biggest challenges reported by the community regarding secondary and tertiary studies process is the protocol elaboration. %
This is the most important item of a secondary or tertiary study, because it documents the executed process aiming to guarantee the success of the study and its replicability \cite{Kitchenham07}. This becomes more difficult when the secondary or tertiary study is conducted by novice researchers, due to their limited experience in the conduct of these type of studies and with lack of familiarity with the investigated research topic \cite{Riaz10}. A lack of knowledge and experience in the conduction of secondary and tertiary studies may end up in biased results. Not less important, conduct secondary and tertiary studies is considered a quite large task and labour intensive \cite{Kitchenham15}. 

Motivated by the complexity of conducting secondary and tertiary studies process and the protocol development challenge, in this paper we propose a Secondary and Tertiary Study Canvas. A canvas is a visual tool that improves the communication about a project or process facilitating the understating of its steps; it also stimulates the collaboration among all people involved. 
Thus, our main objective is to guide SE researchers and practitioners,  working as good practice steps to be considered during the conduction of secondary and tertiary studies.  To this end, we synthesized and organized the existing secondary studies' protocols in a Canvas format, and suggest a step-based approach to assist the secondary and tertiary studies conduction. 

The remainder of this paper is organized as follows. Section \ref{sec:background} provides a brief background on secondary and tertiary studies; Section  \ref{sec:canvas} presents and details the proposed Secondary and Tertiary Studies Canvas; finally, Section \ref{sec:conclusions} concludes our paper and presents future work. 

\section{Background}\label{sec:background}

In this Section, we present a brief background about the secondary and tertiary study process. 


Secondary studies follow a systematic process in which their input are primary studies (e.g., controlled experiments, case studies, surveys) \cite{Kitchenham07}. 
There are two different types of secondary studies in SE: Systematic Literature Review (SLR), also known as Systematic Review (SR), and Systematic Mapping (SM) \cite{Kitchenham07,Petersen15}. The main difference between SLR and SM is that a SM is a more open form of SLR which focuses on provide a broader overview of a topic of interest \cite{Petersen15}. 

The secondary study process involves three main phases: (i) planning, (ii) execution and (iii) reporting the review \cite{Kitchenham15}. These three phases can be conducted interactively; for example, during the execution, a need to modify the protocol can be identified, and then the process returns to the planning phase. In the following, we provide a summary of each phase of the process according to \textit{Kitchenham et al.}'s book \cite{Kitchenham15}. 

\noindent\textbf{(i) Planning phase:} This phase consists of the identification of the need for a secondary study and the protocol development \& validation. If other secondary studies already exist on the same topic and, if these studies already answered the same research questions, there is no need to perform a new secondary study. When the study is outdated, a secondary study update should be performed. Next, the study protocol should be developed.
The protocol is an essential element for secondary studies because it documents the process and its steps that will be executed in the study, including research questions, search strategy to detect relevant studies, study selection criteria, study quality assessment, data extraction and data synthesis process, and reporting the study. When the protocol is finished, it must be validated. The validation can be performed through a pilot test \cite{Kitchenham07} aiming to verify if any modifications or refinements are necessary.

\noindent\textbf{(ii) Execution phase:} After the protocol validation, the execution phase starts. It consists of executing the plan proposed in the study protocol. In summary, in the execution phase the proposed search strategy is executed, then the studies are selected (selection and quality criteria are applied), and then all necessary data to answer the study research questions are extracted and synthesized. 

\noindent\textbf{(iii) Reporting the review phase:} After the execution of the protocol, with all data synthesized and analyzed, it is time to report the review. This phase focuses on the writing of the results to answer the proposed research questions, as well as the dissemination of the results in order to reach potential interested researchers and practitioners. 


Due to a large number of secondary studies available, it became possible to conduct tertiary studies. They follow exactly the same process of the secondary studies already described in this Section. The difference is that instead of summarizing data from primary studies, they consider the summary of secondary studies \cite{Kitchenham07}.

\section{A Secondary and Tertiary Study Canvas}
\label{sec:canvas}

A Canvas is a visual tool already applied in the administration field. The Business Model Canvas proposed by \textit{Osterwalder \& Pigneur} \cite{osterwalder2010business} is largely recognized as a useful tool to develop and document business model. Since this visual tool helps to improve  communication about a project or process, clarifying and facilitating the understanding of a process steps, we opted to bring this idea to EBSE context. Besides of that, the Canvas can contribute to collaboration among secondary and tertiary studies author's due to the fact that its objectiveness and visual characteristics enable to visualize what step of the secondary and tertiary study process is done, in execution and to do yet; contributing to the study tracking and consequently, influencing in the collaboration among the authors. 

Our Secondary and Tertiary Study Canvas is presented in Figure \ref{sec:canvas}. For didactic reasons, we opted to present the Canvas filled with an example of a secondary study already published. The study considered as an example was the systematic mapping conducted by \textit{Di Francesco et al.} \cite{DIFRANCESCO2019}. 
\textbf{The fillable version of the Canvas is available on:\textit{ \url{https://bit.ly/3dLFSN6}}}.

The Canvas starts with a header where the researcher first fills the research area or topic of interest which the study will be performed on, the involved authors' names and the study type. It is fundamental to clarify the study type because it determines what will be the type of studies considered as input (SLR and SM consider primary studies and Tertiary Study considers SLRs and SMs as input) \cite{Kitchenham15}. This fact directly affects the search string construction as well as the selection criteria steps. 

The review  planning starts with identifying the need of the study (see Section \ref{sec:background} -- item (i)). This point could impact the study type chosen; for example, instead of conducting a tertiary study, the researchers conduct a secondary because is not enough secondary evidence available \cite{Kitchenham15}. Next, it is time to define the study objective \textbf{(step 1)}. It consists of clarifying and summarizing what is the study goal. With the study objective ready, it is time to translate this objective into Research Questions (RQs) \textbf{(step 2)}. A secondary or tertiary study can contain one or more RQs, and they can be broader (general appeared in SMs) or narrower (general appeared in SLRs).

The next step on the Canvas is to define a search strategy \textbf{(step 3)}. The search strategy aims to detect the most relevant studies possible (high recall) with less false positive possible (low precision)\cite{Dieste09}. There are three types of search strategy for secondary and tertiary studies. They are:

\begin{inparaenum}[(i)]
\item Automated search -- It consists of searching relevant studies in Digital Libraries (such as ACM Digital Library, IEEE Xplore, or Scopus) using a search string built based on the RQs. The search string must be tested to evaluate its capacity of return relevant studies; for this the creation of a study control group must be considered \cite{Kitchenham15}.
\item Snowballing -- A technique proposed by \textit{Wholim} \cite{Wholin14}, which consists of looking through studies references (``backward snowballing'') and/or the studies citations (``forward Snowballing''). Both snowballing types require a study seed set, which is a group of the selected studies from other search strategies. This point leads to the mandatory execution of another search strategy first. Finally, 
\item Manual search -- It consists of manually searching for relevant studies in specific conferences and journals related to the chosen research topic \cite{Kitchenham15}. 
\end{inparaenum}

Next, the selection criteria \textbf{(step 4)} addresses the creation of inclusion and exclusion criteria that, when applied to the studies returned by the search strategy execution, filter only the relevant studies to answer the RQs. Study Quality criteria also can be created; their focus is to evaluate the quality of the included studies to avoid the inclusion of weak evidence. Quality criteria assessment has been more applied in the context of SLR \cite{Napoleao17}. 

It is important to highlight that all steps described in the planning phase are iterative and incremental: during the execution of any step, one can backtrack by one or more steps.
The protocol must be validated by an experienced research in order to ensure that all planned steps are correct, including: verify if the search string is appropriately derived from the RQs; and if the data extracted from the studies and its analysis accurately answer the RQs. With the protocol validated, the next step is to execute the planned search strategy proposed on step 3 \textbf{(step 5)}. Next, the selection criteria need to be applied in the candidate studies returned by the execution of the search strategy \textbf{(step 6)}. As a result, a final list of included studies is generated.  

The next step \textbf{(step 7)} is the creation of a data extraction form to extract data from the included studies. A data extraction form  i generally a spreadsheet containing fields derived from the RQs. For example, Figure \ref{sec:canvas} describes the RQ: ``What is the focus of research in architecting with microservices?'' Consequently, the data extraction form should contain a field called ``research topics in architecting with microservices''. 

The steps 8 and 9 are dependent and interactive because the data extracted from the included studies depends on the data synthesis method adopted. 
All data extracted \textbf{(step 8)} should be validated and disagreements among reviewers about extra information should be listed and resolved by consensus \cite{Kitchenham07}.
The data synthesis \textbf{(step 9)} can be done in a descriptive way (qualitative), but a quantitative summary can be used to complement the descriptive synthesis. 

After the data is synthesized, the next steps \textbf{(steps 10 to 13)} address reporting the review. In this phase, it is fundamental to define the format and type of report that the study will be reported, as well as the dissemination strategy to reach potentially interested readers.  
The steps 10 to 13 suggested by us are similar to the items \textit{Results, Discussion, Conclusion} from the PRISMA structure discussed by \textit{Kitchenham et al.} \cite{Kitchenham15}. 
Section \textit{Results} \textbf{(step 10)} must report the study's results, including the study selection process (preferably in a diagram format), the data synthesis analysis including graphs and tables if applicable, and the most important: the answers to RQs. 
The section \textit{Discussion} \textbf{(step 11)} must provide a summary of each RQs as well as discuss their findings. In addition, all the limitations of the study must be discussed \cite{Kitchenham15}. 

The \textit{Conclusions} Section \textbf{(step 12)} should provide an interpreted summary of the results presented and discussed in the secondary or tertiary studies. It can also provide recommendations about the addressed topic and future work \cite{Kitchenham15}.

Finally, when the secondary or tertiary report is ready, it must be reviewed carefully by all authors \cite{Kitchenham15} and then submitted following the chosen dissemination strategy \textbf{(step 13)}. 

The presented Secondary and Tertiary Studies Canvas proposed in this paper summarize in a high-level visual way the steps of the secondary and tertiary studies process. It is worth highlighting that the presented steps and descriptions are only good practices translated from the renowned guidelines for tertiary and secondary studies in SE area. Not all steps mentioned are mandatory: for example, it is possible to just choose automated search as a search strategy, instead of combining search strategies; or the study can use only qualitative analysis instead of quantitative and qualitative; or even do not apply study quality criteria as selection criteria.

\section{Conclusions} 
\label{sec:conclusions}

In this paper, we proposed a Secondary and Tertiary Study Canvas. The main contribution of this paper is the Canvas itself and the step-based approach description of the secondary and tertiary studies process provided. We expect the proposed Canvas 
help  researchers  and  practitioners  during  the  entire  conduction  of their  secondary  and  tertiary  studies,  working as good practice steps guide, especially  for novice researchers who generally have more difficulty during the  elaboration of the protocol. 

In addition, we expect that the Canvas will stimulate the collaboration among the authors of secondary and tertiary studies  during  the conduction of their  study.  

As future work, we intend to perform a practical validation of the proposed Canvas considering researchers and practitioners with different levels of experience in the conduction of SLRs, SMs and tertiary studies. 

\begin{figure*}
	\centering
	\includegraphics[width=0.99\textwidth]{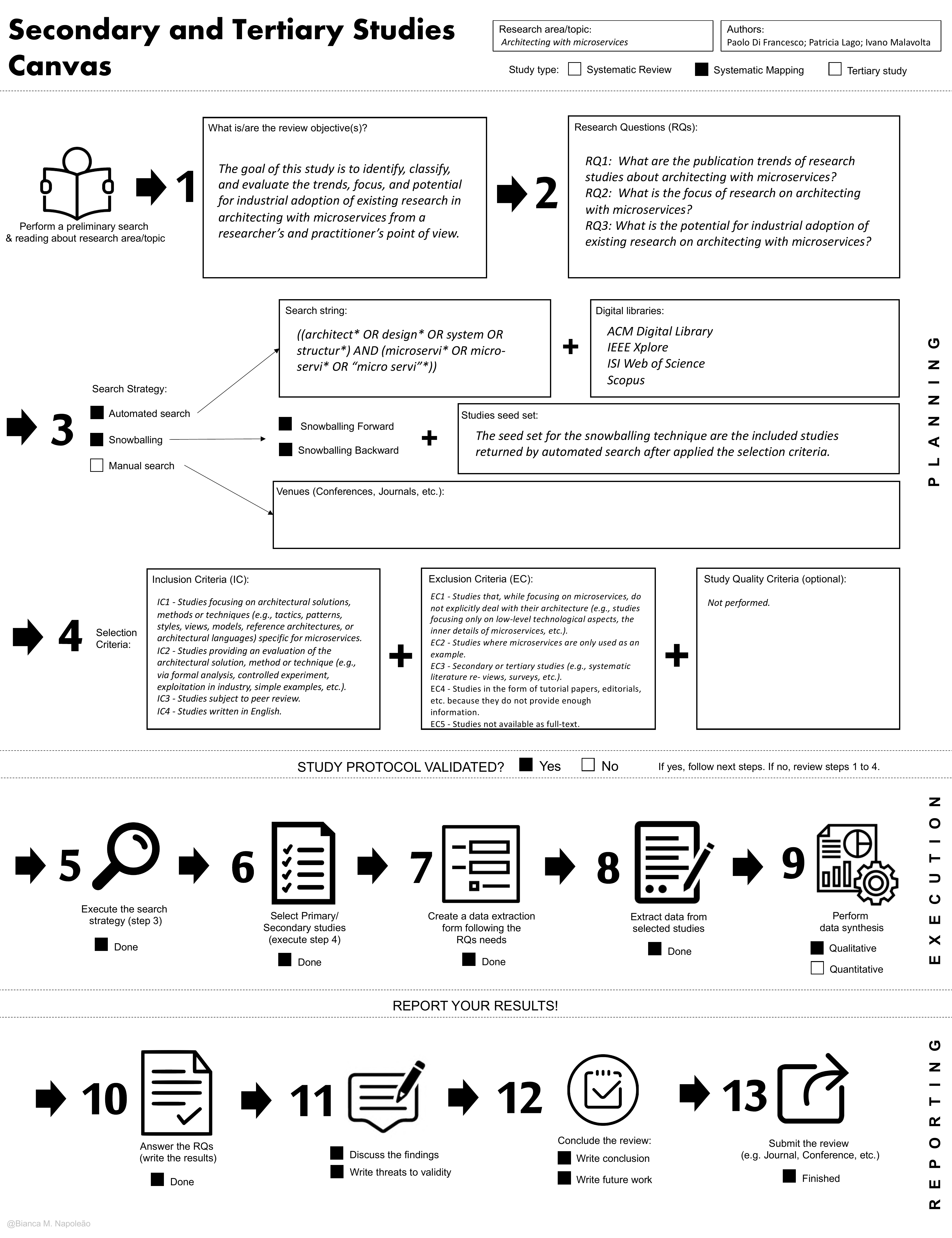}
	\caption{Secondary and Tertiary Studies Canvas (filled with a SM example \cite{DIFRANCESCO2019}).}
	\label{fig:canvas}
\end{figure*}

\bibliographystyle{unsrt}  
\bibliography{references}  

\end{document}